\documentclass{jetpl}
\usepackage{epsf}
\usepackage{graphicx}

\twocolumn

\lat

\title{Testing neutrino magnetic moment in ionization of atoms by
neutrino impact}

\rtitle{Testing neutrino magnetic moment\ldots}

\sodtitle{Testing neutrino magnetic moment in ionization of atoms
by neutrino impact}

\author{K.\,A.\,Kouzakov$^{+}$, A.\,I.\,Studenikin$^{*}$,
M.\,B.\,Voloshin$^{\dag\S}$\/\thanks{e-mail: voloshin@umn.edu}}

\rauthor{K.\,A.\,Kouzakov, A.\,I.\,Studenikin, M.\,B.\,Voloshin}

\sodauthor{Kouzakov, Studenikin, Voloshin}

\address{$^+$Department of Nuclear Physics and Quantum
Theory of Collisions, Faculty of Physics, Moscow State University, Moscow 119991, Russia\\~\\
$^*$Department of Theoretical Physics, Faculty of Physics, Moscow
State University, Moscow 119991, Russia\\~\\
$^\dag$William I. Fine Theoretical Physics Institute, University
of
Minnesota, Minneapolis, Minnesota 55455, USA\\~\\
$^\S$Institute of Theoretical and Experimental Physics, Moscow
117218, Russia}

\dates{\today}{*}

\abstract{The atomic ionization processes induced by scattering of
neutrinos play key roles in the experimental searches for a
neutrino magnetic moment. Current experiments with reactor
(anti)neutrinos employ germanium detectors having energy threshold
comparable to typical binding energies of atomic electrons, which
fact must be taken into account in the interpretation of the data.
Our theoretical analysis shows that the so-called stepping
approximation to the neutrino-impact ionization is well applicable
for the lowest bound Coulomb states, and it becomes exact in the
semiclassical limit. Numerical evidence is presented using the
Thomas-Fermi model for the germanium atom.}


\begin{document}

\maketitle

The neutrino magnetic moments (NMM) expected in the Standard Model
are very small and proportional to the neutrino masses~\cite{fs}:
$\mu_\nu \approx 3 \times 10^{-19}\mu_B(m_\nu/1\,{\rm eV})$ with
$\mu_B = e/2m$ being the electron Bohr magneton, and $m$ is the
electron mass. Thus any larger value of $\mu_\nu$ can arise only
from physics beyond the Standard Model (a recent review of this
subject can be found in Ref.~\cite{gs}). Current direct
experimental searches~\cite{tx,ge1,ge2} for a magnetic moment of
the electron (anti)neutrinos from reactors have lowered the upper
limit on $\mu_\nu$ down to $\mu_\nu < 3.2 \times
10^{-11}\mu_B$~\cite{ge2}. These ultra low background experiments
use germanium crystal detectors exposed to the neutrino flux from
a reactor and search for scattering events by measuring the energy
$T$ deposited by the neutrino scattering in the detector. The
sensitivity of such a search to NMM crucially depends on lowering
the threshold for the energy transfer $T$, due to the enhancement
of the magnetic scattering relative to the standard electroweak
one at low $T$. Namely, the differential cross section $d
\sigma/dT$ is given by the incoherent sum of the magnetic and the
standard cross section, and for the scattering on free electrons
the NMM contribution is given by the formula~\cite{dn,ve}
\begin{eqnarray} \frac{d\sigma_{(\mu)}}{dT }= 4 \pi \alpha
\mu_\nu^2 \frac{1}{T}\left ( 1 - {T \over E_\nu } \right ),
\label{fe} \end{eqnarray}
where $E_\nu$ is the energy of the incident neutrino, and displays
a $1/T$ enhancement at low energy transfer. The standard
electroweak contribution is constant in $T$ at $E_\nu \gg T$:
\begin{eqnarray} \frac{d \sigma_{EW}}{ dT }= {G_F^2 m \over 2 \pi} \left (
1+ 4  \sin^2 \theta_W + 8 \sin^4 \theta_W \right ) \times
\nonumber\\\times \left [ 1 + O \left ( {T \over E_\nu} \right)
\right ] \approx 10^{-47} \, \frac{\rm cm^2 }{\rm keV}.
\label{sew}
\end{eqnarray}
In what follows we refer to these two types of contribution to the
scattering as, respectively, the magnetic and the weak.

The current experiments have reached threshold values of $T$ as
low as few keV and are likely to further improve the sensitivity
to low energy deposition in the detector. At low energies however
one can expect a modification of the free-electron formulas
(\ref{fe}) and (\ref{sew}) due to the binding of electrons in the
germanium atoms, where e.g. the energy of the $K_\alpha$ line,
9.89\,keV, indicates that at least some of the atomic binding
energies are comparable to the already relevant to the experiment
values of $T$. Thus, a proper treatment of the atomic effects in
neutrino scattering is necessary and important for the analysis of
the current and even more of the future data with a still lower
threshold. For the first time this problem was addressed in
Ref.~\cite{gdt75}, where a 2-3 times enhancement of the
electroweak cross section in the case of ionization from a 1$s$
state of a hydrogen-like atom with nuclear charge $Z$ had been
numerically determined at neutrino energies $E_\nu\sim\alpha
Zmc^2$. Subsequent numerical calculations within the
Hartree-Fock-Dirac method for ionization from inner shells of
various atoms showed much lower enhancement ($\sim5-10\%$) of the
electroweak contribution~\cite{ddf92,kmsf,fms}. The interest to
the role of atomic effects was renewed in several recent papers,
which however are ridden by a `trial and error' approach. The
early claim~\cite{wll} of a significant enhancement of the NMM
contribution by the atomic effects has been later
disproved~\cite{mv,wll2} and it was argued~\cite{mv,ks} that the
modification of the formulas (\ref{fe}) and (\ref{sew}) by the
atomic binding effects is insignificant down to very low values of
$T$. It has been subsequently pointed out~\cite{ks} that the
analysis of Ref.~\cite{mv} is generally invalidated in
multi-electron systems, including atoms with $Z > 1$. Furthermore,
the analysis of Ref.~\cite{mv} is also generally invalidated by
singularities of the relevant correlation function in the complex
plane of momentum transfer\footnote{The flaws in the
momentum-transfer dispersion relation and sum rules of
Ref.~\cite{mv} are corrected in Ref.~\cite{ksv}.}, so that the
claimed behavior of the cross section at low $T$ applies only in
the semiclassical limit.

In this paper we revisit the subject of neutrino scattering on
atoms at low energy transfer. We aim at describing this process at
$T$ in the range of few keV and lower, so that the motion of the
electrons is considered as strictly nonrelativistic. Also in this
range the energy of the dominant part of the incident neutrinos
from the reactor is much larger than $T$ and we thus neglect any
terms whose relative value is proportional to $T/E_\nu$.
Furthermore  any recoil of the germanium atom as a whole results
in an energy transfer less than $2E_\nu^2/M_{Ge}$, which at the
typical reactor neutrino energy is well below the considered here
keV range of the energy transfer. Thus we formally set the mass of
the atomic nucleus to infinity and  neglect any recoil by the atom
as a whole.  In particular, under these conditions the interaction
of the neutrino with the nucleus can be entirely neglected, and
only the scattering on the atomic electrons is to be considered.

The kinematics of the scattering of a neutrino on atomic electrons
is generally characterized by the components of the four-momentum
transfer, the energy transfer $T$ and the spatial momentum
transfer ${\bf q}$, from the neutrino to the electrons with two
rotationally invariant variables being $T$ and $q=|{\bf q}|$. At
small $T$ the electrons can be treated nonrelativistically both in
the initial and the final state, so that the process is that of
scattering of an NMM in the electromagnetic field $A=(A_0,{\bf
A})$ of the electrons: $A_0({\bf q})=\sqrt{4 \pi \alpha}\rho( {\bf
q})/ q^2$, ${\bf A}({\bf q})=\sqrt{4\pi\alpha}{\bf j}({\bf
q})/q^2$, where $\rho({\bf q})$ and ${\bf j}({\bf q})$ are the
Fourier transforms of the electron number density and current
density operators, respectively,
\begin{eqnarray} \rho({\bf q})&=& \sum_{a=1}^Z \exp(i {\bf qr}_a),
\label{ne}\\
{\bf j}({\bf q})&=& -\frac{i}{2m}\sum_{a=1}^Z \left[\exp(i {\bf
qr}_a)\frac{\partial}{\partial{\bf
r}_a}+\frac{\partial}{\partial{\bf r}_a}\exp(i {\bf qr}_a)\right],
\label{ce} \end{eqnarray}
and the sums run over the positions ${\bf r}_a$ of all the $Z$
electrons in the atom.

In this limit the expression for the double differential cross
section is given by~\cite{ks}
\begin{eqnarray} \frac{d^2 \sigma_{(\mu)} }{ dT d q^2} = 4 \pi  \alpha
\frac{ \mu_\nu^2 }{ q^2}
\left[\left(1-\frac{T^2}{q^2}\right)S(T,q^2)+\right.\nonumber\\+\left.\left(1-\frac{q^2}{4E_\nu^2}\right)R(T,q^2)\right],
\label{d2s} \end{eqnarray}
where $S(T,q^2)$, also known as the dynamical structure
factor~\cite{vh}, and $R(T,q^2)$ are
\begin{eqnarray} S(T,q^2)&=&\sum_n \delta (T - E_n+E_0) \left | \langle n
| \rho({\bf q}) | 0 \rangle \right |^2, \label{dsf} \\
 R(T,q^2)&=&\sum_n \delta (T - E_n+E_0) \left | \langle n |
j_\perp({\bf q}) | 0 \rangle \right |^2, \label{dsfc}
\end{eqnarray}
with $j_\perp$ being the ${\bf j}$ component perpendicular to
${\bf q}$ and parallel to the scattering plane, which is formed by
the incident and final neutrino momenta. The sums in
Eqs.~(\ref{dsf}) and~(\ref{dsfc}) run over all the states $| n
\rangle$ with energies $E_n$ of the electron system, with $|0
\rangle$ being the initial state.

Clearly, the factors $S(T,q^2)$ and $R(T,q^2)$ are related to
respectively the density-density and current-current Green's
functions
\begin{eqnarray} F(T,q^2)&=&\sum_n \frac{\left | \langle n | \rho({\bf q}) | 0
\rangle \right |^2 }{ T - E_n+E_0 - i \epsilon} \nonumber \\&=&
\left \langle 0 \left |\rho(- {\bf q}) \frac{1 }{ T-H+E_0- i
\epsilon}\rho({\bf q}) \right | 0 \right \rangle, \label{fdef}\\
L(T,q^2)&=&\sum_n \frac{\left | \langle n | j_\perp({\bf q}) | 0
\rangle \right |^2 }{ T - E_n+E_0 - i \epsilon} \nonumber\\ &=&
\left \langle 0 \left |j_\perp(- {\bf q}) \frac{1 }{ T-H+E_0- i
\epsilon} j_\perp({\bf q}) \right | 0 \right \rangle,
\label{fdefc}
\end{eqnarray}
as
\begin{equation} S(T,q^2)=\frac{1 }{ \pi} \Imag F(T,q^2), \label{sfrel} \end{equation}
\begin{equation} R(T,q^2)=\frac{1 }{ \pi} \Imag L(T,q^2), \label{sfrelc} \end{equation}
with $H$ being the Hamiltonian for the system of electrons. For
small values of $q$, in particular, such that $q\sim T$, only the
lowest-order non-zero terms of the expansion of Eqs.~(\ref{sfrel})
and~(\ref{sfrelc}) in powers of $q^2$ are of relevance (the
so-called dipole approximation). In this case, one has~\cite{ks}
\begin{equation} R(T,q^2)=\frac{T^2}{q^2}S(T,q^2). \label{da} \end{equation}

Taking into account Eq.~(\ref{da}), the experimentally measured
single-differential inclusive cross section is, to a good
approximation, given by (see e.g. in Refs.~\cite{mv,ks})
\begin{equation} \frac{d \sigma_{(\mu)} }{ dT } = 4 \pi  \alpha \mu_\nu^2
\int_{T^2}^{4E_\nu^2} S(T,q^2) \frac{dq^2 }{ q^2}. \label{d1s}
\end{equation}

The standard electroweak contribution to the cross section can be
similarly expressed in terms of the same factor
$S(T,q^2)$~\cite{mv} as
\begin{eqnarray} \frac{d \sigma_{EW} }{ dT } = \frac{G_F^2 }{ 4 \pi} \left ( 1+ 4 \sin^2 \theta_W + 8 \sin^4 \theta_W \right ) \times \nonumber \\
\times \int_{T^2}^{4E_\nu^2} S(T,q^2) dq^2 , \label{d1sw}
\end{eqnarray}
where the factor $S(T,q^2)$ is integrated over $q^2$ with a unit
weight, rather than $q^{-2}$ as in Eq.~(\ref{d1s}).

The kinematical limits for $q^2$ in an actual neutrino scattering
are explicitly indicated in Eqs.~(\ref{d1s}) and (\ref{d1sw}). At
large $E_\nu$, typical for the reactor neutrinos, the upper limit
can in fact be extended to infinity, since in the discussed here
nonrelativistic limit the range of momenta $\sim E_\nu$ is
indistinguishable from infinity.  The lower limit can be  shifted
to $q^2=0$, since the contribution of the region of $q^2 < T^2$
can be expressed in terms of the photoelectric cross
section~\cite{mv} and is negligibly small (at the level of below
one percent in the considered range of $T$). For this reason we
henceforth discuss the momentum-transfer integrals in
Eqs.~(\ref{d1s}) and~(\ref{d1sw}) running from $q^2=0$ to
$q^2=\infty$:
\begin{subequations}\label{defi}
\begin{align}
I_1(T)&=\int_0^{\infty} S(T,q^2) \frac{dq^2 }{ q^2},\\
 I_2(T)&=\int_0^{\infty} S(T,q^2)\,dq^2.
 \end{align}
\end{subequations}

For a free electron, which is initially at rest, the
density-density correlator is the free particle Green's function
\begin{equation} F_{(FE)}(T,q^2)= \left ( T-\frac{q^2 }{ 2m} - i \epsilon
\right )^{-1} \label{ff} \end{equation}
so that the dynamical structure factor is given by
$S_{(FE)}(T,q^2)=\delta(T-q^2/2m)$, and the discussed here
integrals are in the free-electron limit as follows:
\begin{subequations} \label{intf}
\begin{align} I_1^{(FE)}&=\int_0^{\infty} S_{(FE)}(T,q^2)\frac{dq^2
}{ q^2} = \frac{1 }{ T}, \\ I_2^{(FE)}&=\int_0^{\infty}
S_{(FE)}(T,q^2)\, dq^2 = 2 m.
\end{align}
\end{subequations}
It is readily seen that these expressions, when used in the
formulas (\ref{d1s}) and (\ref{d1sw}), result in the free-electron
cross section in Eqs.~(\ref{fe}) and~(\ref{sew}).

Let us consider the scattering on just one bound electron. The
Hamiltonian for the electron has the form $H=p^2/2m + V(r)$, and
the density-density Green's function from Eq.~(\ref{fdef}) can be
written as
\begin{eqnarray}
F(T,q^2)&=& \left \langle 0 \left | e^{-i {\bf qr}} \left [ T -H + E_0 \right ]^{-1}  e^{i {\bf qr}}\right | 0 \right \rangle \nonumber \\
&=&
 \left \langle 0 \left |  \left [ T -\frac{q^{2} }{ 2 m}- \frac{{\bf pq} }{ m}  - H + E_0  \right ]^{-1} \right | 0 \right \rangle
 ,\nonumber\\
\label{f1}
\end{eqnarray}
where the infinitesimal shift $T \to T - i \epsilon$ is implied.

Clearly, a nontrivial behavior of the latter expression in
Eq.~(\ref{f1}) is generated by the presence of the operator ${\bf
pq}/m$ in the denominator, and the fact that it does not commute
with the Hamiltonian $H$. Thus an analytical calculation of the
Green's function as well as the dynamical structure factor and the
momentum-transfer integrals is feasible in only few specific
problems. In particular, such a calculation for ionization from
the $1s$, $2s$, and $2p$ hydrogen-like states shows that the
deviation of the discussed integrals (\ref{defi}) from their free
values are very small~\cite{ksv}: the largest deviation is exactly
at the ionization threshold, where, for instance, each of the $1s$
integrals is equal to the free-electron value multiplied by the
factor $(1-7 e^{-4}/3) \approx 0.957$, and in the $2s$ and $2p$
cases the departure from the free-electron behavior is even
smaller.

The problem of calculating the integrals (\ref{defi}) however can
be solved in the semiclassical limit, where one can neglect the
noncommutativity of the momentum ${\bf p}$ with the Hamiltonian,
and rather treat this operator as a number vector. Taking also
into account that $(H-E_0) |0 \rangle =0$, one can then readily
average the latter expression in Eq.~(\ref{f1}) over the
directions of ${\bf q}$ and find the formula for the dynamical
structure factor:
\begin{eqnarray} S(T,q^2)=\frac{m }{ 2  p  q} \left [ \theta \left ( T- \frac{q^2
}{ 2m}+\frac{p q }{ m} \right) -\right.\nonumber\\ \left. -\theta
\left ( T- \frac{q^2 }{ 2m}- \frac{p q }{ m} \right) \right ],
\label{scls} \end{eqnarray}
where $p= |{\bf p}|$ and $\theta$ is the standard Heaviside step
function. The expression in Eq.~(\ref{scls}) is nonzero only in
the range of $q$ satisfying the condition $-pq/m < T - q^2/2m <
pq/m$, i.e. between the (positive) roots of the binomials in the
arguments of the step functions: $q_{min}=\sqrt{2m T + p^2} - p$
and $q_{max}=\sqrt{2m T + p^2} + p$. One can notice that the
previously mentioned `spread and shift' of the peak in the
dynamical structure function in this limit corresponds to a flat
pedestal between $q_{min}$ and $q_{max}$. The calculation of the
integrals (\ref{defi}) with the expression (\ref{scls}) is
straightforward, and yields the free-electron expressions
(\ref{intf}) for the discussed here integrals in the semiclassical
(WKB) limit:
\begin{equation} I_1^{(WKB)}=\frac{1 }{ T}, \qquad I_2^{(WKB)}=2 m.
\label{wkbi} \end{equation}
The difference from the pure free-electron case however is in the
range of the energy transfer $T$. Namely, the expressions
(\ref{wkbi}) are applicable in this case only above the ionization
threshold, i.e. at $T \ge |E_0|$. Below the threshold the electron
becomes `inactive'.

It is instructive to point out that the validity of the  result in
Eq.~(\ref{wkbi}) is based on the semiclassical approximation and
is not directly related to the value of the energy $T$. In
particular, for a Coulomb interaction the WKB approximation is
applicable at energy near the threshold~\cite{ll}. For $T$ exactly
at the threshold, $T=-E_0$, the criterion for applicability of the
semiclassical approach in terms of the force $F=|{\bf F}| = |[{\bf
p}, H]|$ acting on the electron and the momentum $p$ of the
electron is that~\cite{ll} the ratio of the characteristic values
$m F/p^3$ is small. For the excitation of a state with the
principal number $n$  one has $|F|=\alpha/r^2 \sim m^2 \alpha^3
n^4$ and $p \sim m \alpha/n$, so that $m |F|/p^3 \sim 1/n$. Thus
the applicability of a semiclassical treatment of the ionization
near the threshold improves for initial states with large $n$. As
previously mentioned, the modification of the integrals
(\ref{defi}) by the binding is already less than 5\% for $n=1$, so
that we fully expect this deviation to be  smaller for the higher
states, and even smaller at larger values of $T$ above the
threshold due to the approach to the free-electron behavior at $T
\gg E_0$.

We believe that the latter conclusion explains the so-called
stepping behavior observed empirically~\cite{kmsf} in the results
of numerical calculations. Namely the calculated cross section $d
\sigma/dT$ for ionization of an electron from an atomic orbital
follows the free-electron dependence on $T$ all the way down to
the threshold for the corresponding orbital with a very small, at
most a few percent, deviation. This observation led the authors of
Ref.~\cite{kmsf} to suggest the stepping approximation for the
ratio of the atomic cross section (per target electron) to the
free-electron one:
\begin{equation} f(T) \equiv \frac{ d \sigma / dT }{ (d \sigma/dT)_{FE}} =
\frac{1 }{ Z}  \sum_i  n_i  \theta(T - |E_i|), \label{step}
\end{equation}
where the sum runs over the atomic orbitals with the binding
energies $E_i$ and the filling numbers $n_i$. Clearly, the factor
$f(T)$ simply counts the fraction of `active' electrons at the
energy $T$, i.e. those for which the ionization is kinematically
possible. For this reason we refer to $f(t)$ as an
\emph{activation factor}.

\begin{figure}[ht]
\begin{center}
\includegraphics[width=82mm]{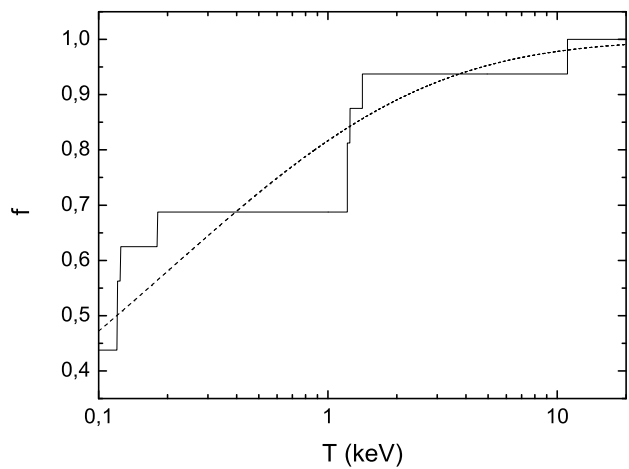}
   \caption{\label{TF_factor}
Fig. 1. The activation factor $f$ for germanium in the stepping
approximation with the actual energies of the orbitals (solid
line) and its value in the Thomas-Fermi model (dashed). }
\end{center}
\end{figure}

In considering the neutrino scattering on actual atoms one needs
to evaluate the dependence of the number of active electrons on
$T$. The energies of the inner $K$, $L$, and $M$ orbitals in the
germanium atom are well known (see e.g. Ref.~\cite{fms} and
references therein) and provide the necessary data for a
description of the neutrino scattering by the stepping formula
(\ref{step}) down to the values of the energy transfer $T$ in the
range of the binding of the $M$ electrons, i.e. at $T \gtrsim
|E_M| \sim 0.1$\,keV. The corresponding steps in the activation
factor are shown in Fig.~\ref{TF_factor}.

Let us show that the stepping approximation~(\ref{step}) can also
be explained in terms of quantum corrections to the activation
factor given by the Thomas-Fermi atomic model (see e.g.
Ref.~\cite{ll}). The latter describes electrons in an atom as a
degenerate free electron gas in a master potential $\phi(r)$
filling the momentum space up to the zero Fermi energy, i.e. up to
the momentum $p_0(r)$ such that $p_0^2/2 m - e \phi=0$. The
electron density $n(r)=p_0^3/(3 \pi^2)$ then determines the
potential $\phi(r)$ from the usual Poisson's equation, thus
resulting in a self-consistent procedure. In the discussed picture
at an energy transfer $T$ the ionization is possible only for the
electrons whose energies in the potential are above $-T$, i.e.
with momenta above $p_T(r)$ with $p_T^2/2m - e \phi = -T$. The
electrons with lower energy are inactive. Calculating the density
of the inactive electrons as $p_T^3/(3 \pi^2)$ and subtracting
their total number from $Z$, one readily arrives at the formula
for the activation factor, i.e. the effective fraction of the
active electrons $Z_{\rm eff}/Z$ as a function of $T$:
\begin{equation} f(T)=\frac{Z_{\rm eff}(T) }{ Z}= 1 - \int_0^{x_0(T)} \left [
\frac{\chi(x) }{ x} - \frac{T }{ T_0} \right ]^{3/2} x^2 dx,
\label{zeff} \end{equation}
where $\chi(x)$ is the Thomas-Fermi function, well known and
tabulated, of the scaling variable $x = 2 (4/3\pi)^{2/3} m \alpha
Z^{1/3}$,  the energy scale $T_0$ is given by
\begin{equation} T_0=2 \left ( \frac{4 }{ 3 \pi} \right )^{2/3} m \alpha^2
Z^{4/3} \approx 30.8 Z^{4/3}\,{\rm eV}, \label{t0} \end{equation}
and, finally, $x_0(T)$ is the point where the integrand becomes
zero, i.e. corresponding to the radius beyond which all the
electrons are active at the given energy $T$. The energy scale
$T_0$ in germanium (Z=32) evaluates to  $T_0 \approx 3.1$\,keV.
The activation factor for germanium calculated from the formula
(\ref{zeff}) is shown by the dashed line in the plot of Fig.~2.
One can see that the stepping activation factor~(\ref{step})
mimics upon average over the energy intervals between the electron
shells in germanium the Thomas-Fermi result. Thus, it can be
considered as refinement of the latter due to account for the
quantization of the electron binding energies.

It should be remarked that the discussed statistical model is
known to approximate the average bulk properties of the atomic
electrons with a relative accuracy $O(Z^{-2/3})$ and as long as
the essential distances $r$ satisfy the condition $Z^{-1} \ll m
\alpha r \ll 1$, which condition in terms of the scaling variable
$x$ reads as $Z^{-2/3} \ll x \ll Z^{1/3}$. In terms of the formula
(\ref{zeff}) for the number of active electrons, the lower bound
on the applicability of the model is formally broken at $T \sim
Z^{2/3} T_0$, i.e. at the energy scale of the inner atomic shells.
However the effect of the deactivation of the inner electrons is
small, of order $Z^{-1}$ in comparison with the total number $Z$
of the electrons. On the other hand, at low $T$, including the
most interesting region of $T \sim T_0$, the integral in
Eq.~(\ref{zeff}) is determined by the range of $x$ of order one,
where the model treatment is reasonably justified.

We have considered the scattering of neutrinos on electrons bound
in atoms. Our main finding is that the differential over the
energy transfer cross section given by the free-electron formulas
(\ref{fe}) and (\ref{sew}) and the stepping behavior of the
activation factor given by Eq.~(\ref{step}) provides a very
accurate description of the neutrino-impact ionization of a
complex atom, such as germanium, down to quite low energy
transfer. The deviation from this approximation due to the onset
of the ionization near the threshold is less than 5\% (of the
height of the step) for the $K$ electrons, if one applies the
analytical behavior of this onset that we find for the ground
state of a hydrogen-like ion. We also find that the free-electron
expressions for the cross section are not affected by the atomic
binding effects in the semiclassical limit. For this reason we
expect that the deviation of the actual onset from a step function
at the threshold for ionization of higher atomic orbitals is even
smaller than for the ground state, since the motion in the higher
states is closer to the semiclassical limit.  Thus, our analytical
results explain the numerically determined behaviors of the
electroweak and magnetic contributions to the neutrino-impact
ionization of various atomic targets~\cite{ddf92,kmsf,fms}.

%
%
We thank A.\,S.~Starostin and Yu.\,V.~Popov for useful and
stimulating discussions.  The work of K.A.K. (in part) and A.I.S.
is supported by RFBR grant 11-02-01509-a. K.A.K. also acknowledges
partial support from RFBR grant 11-01-00523-a. The work of M.B.V.
is supported in part by the DOE grant DE-FG02-94ER40823.
%
%

\end{document}